\documentclass{ieeeaccess}
\usepackage{cite}
\usepackage{amsmath,amssymb,amsfonts}
\usepackage{algorithmic}
\usepackage{graphicx}
\usepackage{textcomp}
\usepackage{booktabs}
\usepackage{threeparttable}
\usepackage{float}
\def\BibTeX{{\rm B\kern-.05em{\sc i\kern-.025em b}\kern-.08em
    T\kern-.1667em\lower.7ex\hbox{E}\kern-.125emX}}

\newdimen\xfigwd
\xfigwd=\linewidth

\begin{document}
\history{Date of publication xxxx 00, 0000, date of current version xxxx 00, 0000.}
\doi{10.1109/TQE.2020.DOI}

\title{\textbf{\textbf{A NISQ-Aware Hybrid Quantum–Classical Framework for Scalable Combinatorial Optimization}}}
\author{\uppercase{Haolong Ding}\authorrefmark{1}, 
\uppercase{Mohan Wu\authorrefmark{1},Yin Xu \authorrefmark{1} and Hua Xu}.\authorrefmark{1,2}}
\address[1]{College of Artificial Intelligence, Tianjin University of Science and Technology, Tianjin 300457, China }
\address[2]{Yiwei Quantum Technology Co., Ltd, Hefei 230088, China}
\address[3]{These authors contributed equally:Haolong Ding, Mohan Wu}

\markboth
{Author \headeretal: Preparation of Papers for IEEE Transactions on Quantum Engineering}
{Author \headeretal: Preparation of Papers for IEEE Transactions on Quantum Engineering}

\corresp{Corresponding author: Hua Xu.}

\begin{abstract}
Scalable combinatorial optimization under resource-constrained quantum hardware remains a fundamental challenge in the Noisy Intermediate-Scale Quantum (NISQ) era, due to the mismatch between exponentially growing solution spaces and limited quantum computational capacity. In this work, we propose a NISQ-aware hybrid quantum–classical optimization framework that reformulates large-scale combinatorial optimization as a resource-bounded distribution evolution process. Instead of directly optimizing individual solutions, the proposed framework operates on a probabilistic representation of the solution space, enabling efficient exploration under hardware constraints. Specifically, large problem instances are decomposed into qubit-compatible subproblems via clustering-based decomposition, ensuring resource-bounded optimization. Within each subproblem, a quantum genetic algorithm evolves the solution distribution, while periodically embedded amplitude amplification acts as a controlled quantum enhancement mechanism that accelerates convergence without increasing circuit depth. A classical refinement stage ensures global solution consistency. Extensive experiments on benchmark and synthetic datasets demonstrate that the proposed framework consistently outperforms classical and quantum-inspired baselines, with performance gains that become more pronounced as problem scale increases. This scale-dependent behavior indicates that scalability is achieved through structured decomposition rather than increased quantum complexity. Noise simulations further confirm robustness under realistic NISQ conditions, and ablation studies validate that both quantum evolutionary search and amplitude amplification contribute significantly to performance improvements. Overall, this work establishes a practical and scalable paradigm for hybrid quantum–classical optimization, demonstrating that effective large-scale optimization can be achieved through structured problem decomposition and controlled quantum enhancement under realistic hardware constraints.
\end{abstract}


\titlepgskip=-15pt

\maketitle

\section{Introduction}
\label{sec:introduction}
Combinatorial optimization over large discrete solution spaces remains a fundamental challenge for both classical and quantum approaches. Among these problems, the Traveling Salesman Problem (TSP) serves as a canonical benchmark, characterized by factorial growth in its solution space and classified as NP-hard\cite{1}. As problem size increases, exact methods quickly become computationally intractable, making heuristic and metaheuristic approaches essential in practice\cite{8}. Classical approaches such as genetic algorithms (GAs) and ant colony optimization (ACO) have demonstrated strong performance on small- to medium-scale instances; however, their effectiveness often degrades in high-dimensional search spaces due to premature convergence and limited exploration capability.

In parallel, quantum algorithms offer potential advantages in search efficiency. However, their practical deployment in the Noisy Intermediate-Scale Quantum (NISQ) era is constrained by limited qubit counts, shallow circuit depth, and sensitivity to noise \cite{2}. These limitations fundamentally restrict the direct applicability of quantum algorithms to large-scale combinatorial optimization problems\cite{li2022large}. Recent research has explored hybrid quantum–classical approaches, including quantum-inspired evolutionary algorithms and quantum genetic algorithms (QGA) \cite{4}, which incorporate quantum representations into classical optimization frameworks. While these approaches enhance exploration capability , they remain limited in scalability when applied to large problem instances under realistic NISQ constraints \cite{5}. Existing approaches either rely on classical decomposition without effectively leveraging quantum-enhanced search mechanisms\cite{6}, or apply quantum techniques in a manner that does not scale to large problem instances under practical quantum hardware constraints\cite{tomesh2023divide}.

A central challenge, therefore, is how to design hybrid quantum–classical optimization frameworks that can scale to large problem sizes while remaining compatible with NISQ hardware limitations\cite{osaba2024hybrid}. From a fundamental perspective, this challenge arises from a structural mismatch between the exponential growth of the solution space and the limited representational and computational capacity of near-term quantum devices\cite{bharti2022noisy}. Effective solutions must therefore simultaneously address two tightly coupled problems:
\begin{enumerate}
\item how to restructure the global optimization problem into resource-compatible subspaces.

\item how to perform efficient search within these constrained subspaces.
\end{enumerate}

To address this challenge, we propose a NISQ-aware hybrid quantum–classical optimization framework that reformulates large-scale combinatorial optimization into a structured, resource-bounded optimization paradigm. Instead of directly applying quantum algorithms to the global problem, the proposed framework decomposes the problem into qubit-compatible subproblems, performs localized probabilistic search with quantum enhancement within each subproblem, and reconstructs a globally consistent solution through classical refinement. Within this framework, a quantum genetic algorithm is employed to evolve a probabilistic representation of candidate solutions, enabling distribution-level exploration under hardware constraints, while periodically embedded amplitude amplification provides controlled quantum enhancement that accelerates convergence without increasing circuit depth. This design enables a balanced integration of problem decomposition, probabilistic search, and quantum enhancement under realistic NISQ constraints\cite{perelshtein2023nisq}.

The main contributions of this work are as follows:
\begin{enumerate}
\item We propose a resource-bounded hybrid optimization paradigm that reformulates large-scale combinatorial optimization into qubit-compatible subproblems, enabling scalable optimization under realistic NISQ constraints.

\item We introduce a Grover-enhanced quantum evolutionary mechanism that integrates amplitude amplification into distribution-based search, providing controlled convergence acceleration without increasing circuit depth.

\item We develop a unified hybrid optimization framework that couples probabilistic quantum search with classical reconstruction and refinement, forming a closed-loop distribution evolution system.

\item We provide comprehensive empirical validation on benchmark and synthetic datasets, demonstrating improved solution quality, scalability, robustness under noise, and clear mechanism-level contributions.
\end{enumerate}

The remainder of this paper is organized as follows. Section II reviews the methodological foundations, including quantum genetic algorithms, amplitude amplification, and clustering-based decomposition. Section III presents the proposed hybrid optimization framework. Section IV provides experimental evaluation, including performance, scalability, noise robustness, and ablation analysis. Section V discusses practical implications, limitations, and generalization potential. Section VI concludes the paper.

Taken together, these contributions provide a structured foundation for designing scalable hybrid quantum–classical optimization methods under NISQ constraints\cite{osaba2024hybrid}.

\section{\textbf{Preliminaries and Methodological Foundations}}
This section introduces the core methodological components underlying the proposed hybrid optimization framework, including the Quantum Genetic Algorithm (QGA), Grover-based amplitude amplification, and K-means clustering. These components are presented as fundamental building blocks and will be systematically integrated in Section III into a unified optimization framework designed for scalability and compatibility with NISQ hardware constraints\cite{12}.

\subsection{\textbf{\textbf{Quantum Genetic Algorithm}}}

The Quantum Genetic Algorithm (QGA) is a hybrid optimization paradigm that combines evolutionary search mechanisms with quantum-inspired probabilistic representations\cite{13}. Unlike classical genetic algorithms that explicitly maintain populations of candidate solutions, QGA represents solution distributions implicitly through quantum states, enabling compact encoding and enhanced exploration of the search space\cite{wei2014survey} .In QGA, candidate solutions are represented by a parameterized quantum state, where probability amplitudes encode a distribution over feasible configurations rather than a single deterministic solution. This probabilistic encoding allows the algorithm to maintain diversity while gradually biasing the search toward high-quality regions of the solution space\cite{9} .

The optimization process in QGA typically consists of three stages:
\begin{enumerate}
\item Initialization: A quantum state is prepared, often corresponding to a uniform superposition over candidate solutions.

\item State update: Parameterized quantum operations (e.g., rotation gates) are applied to adjust probability amplitudes based on feedback from evaluated samples\cite{xiong2018quantum}.

\item Measurement and evaluation: The quantum state is measured to obtain classical candidate solutions, which are then evaluated using a problem-specific fitness function.
\end{enumerate}

Compared to classical evolutionary algorithms, QGA provides a probabilistic search mechanism that enables both exploration and guided exploitation of the solution space\cite{17}. 

\subsection{\textbf{\textbf{Amplitude Amplification via Grover's Algorithm}}}
Grover’s algorithm provides a mechanism for amplifying the probability of marked states through iterative amplitude amplification \cite{grover1996fast}. While originally developed for unstructured search, its underlying principle can be adapted to optimization settings by increasing the sampling probability of high-quality solutions\cite{morales2018variational}.

The algorithm operates by applying a sequence of phase marking and diffusion operations to a quantum state. Candidate solutions satisfying a given condition are marked by an oracle operator, and their amplitudes are subsequently increased relative to other states.

In practical settings, amplitude amplification can be applied under the following constraints:
\begin{enumerate}
\item In practical settings, amplitude amplification can be applied under the following constraints:
\item A limited number of amplification iterations is performed;
\item The procedure is applied selectively rather than continuously.
\end{enumerate}

These properties allow amplitude amplification to increase the likelihood of sampling desirable solutions while maintaining compatibility with limited quantum resources, such as shallow circuit depth\cite{moussa2019function}.

\subsection{\textbf{\textbf{K-means Clustering for Problem Decomposition}}}
K-means clustering is a classical unsupervised learning algorithm that partitions data into clusters by minimizing intra-cluster variance. Given a set of data points, the algorithm assigns each point to the nearest cluster centroid and iteratively updates the centroids to improve clustering quality. \cite{lu2016applying}.

In the context of combinatorial optimization, clustering can be used to partition a large problem instance into smaller subproblems based on structural or spatial properties. For example, in a Traveling Salesman Problem (TSP), cities can be grouped into clusters based on their coordinates. The objective function is defined as:
\begin{equation}
  J = \sum_{i=1}^{K} \sum_{x \in C_i} \| x - \mu_i \|^2,
\end{equation}
where $\mu_i = \frac{1}{|C_i|} \sum_{x \in C_i} x$ is the centroid of cluster $C_i$.

The primary role of clustering in this framework is to transform a large-scale optimization problem into a set of bounded subproblems. By controlling the size of each subcluster, the number of cities per subproblem is limited, which directly constrains the required number of qubits and circuit depth for quantum operations\cite{kovacs2018intraclustsp}. This decomposition provides several advantages:
\begin{enumerate}
\item It reduces the effective search space of each subproblem.
\item It enables parallel or independent optimization across subclusters.
\item It ensures compatibility with practical quantum-resource constraints.
\end{enumerate}

Overall, these components provide complementary capabilities for scalable hybrid optimization: QGA enables probabilistic exploration of the solution space, amplitude amplification provides controlled enhancement of promising search regions, and clustering-based decomposition ensures scalable optimization under practical NISQ resource constraints.

\section{\textbf{Proposed NISQ-aware Hybrid Optimization Framework}}
Building upon the methodological components introduced in Section II, we propose a NISQ-aware hybrid quantum–classical optimization framework for solving large-scale combinatorial optimization problems\cite{magann2022feedback} .

The central design principle is to reformulate a globally intractable problem into a structured, resource-bounded optimization paradigm, in which problem decomposition, probabilistic search, and controlled quantum enhancement are jointly coordinated. Instead of directly applying quantum algorithms to the global problem, the framework operates as a hierarchical hybrid system that decomposes the original problem into qubit-compatible subproblems, performs localized probabilistic search with quantum enhancement, and reconstructs a globally consistent solution through classical refinement\cite{24}.

From a unified perspective, the framework can be interpreted as a distribution-level optimization process under resource constraints, where a probability distribution over candidate solutions is iteratively reshaped through the interaction between quantum state evolution and classical feedback\cite{gilliam2021grover}.

\subsection{\textbf{\textbf{ Framework Overview}}}
\begin{figure*}
    \centering
    \includegraphics[width=\textwidth]{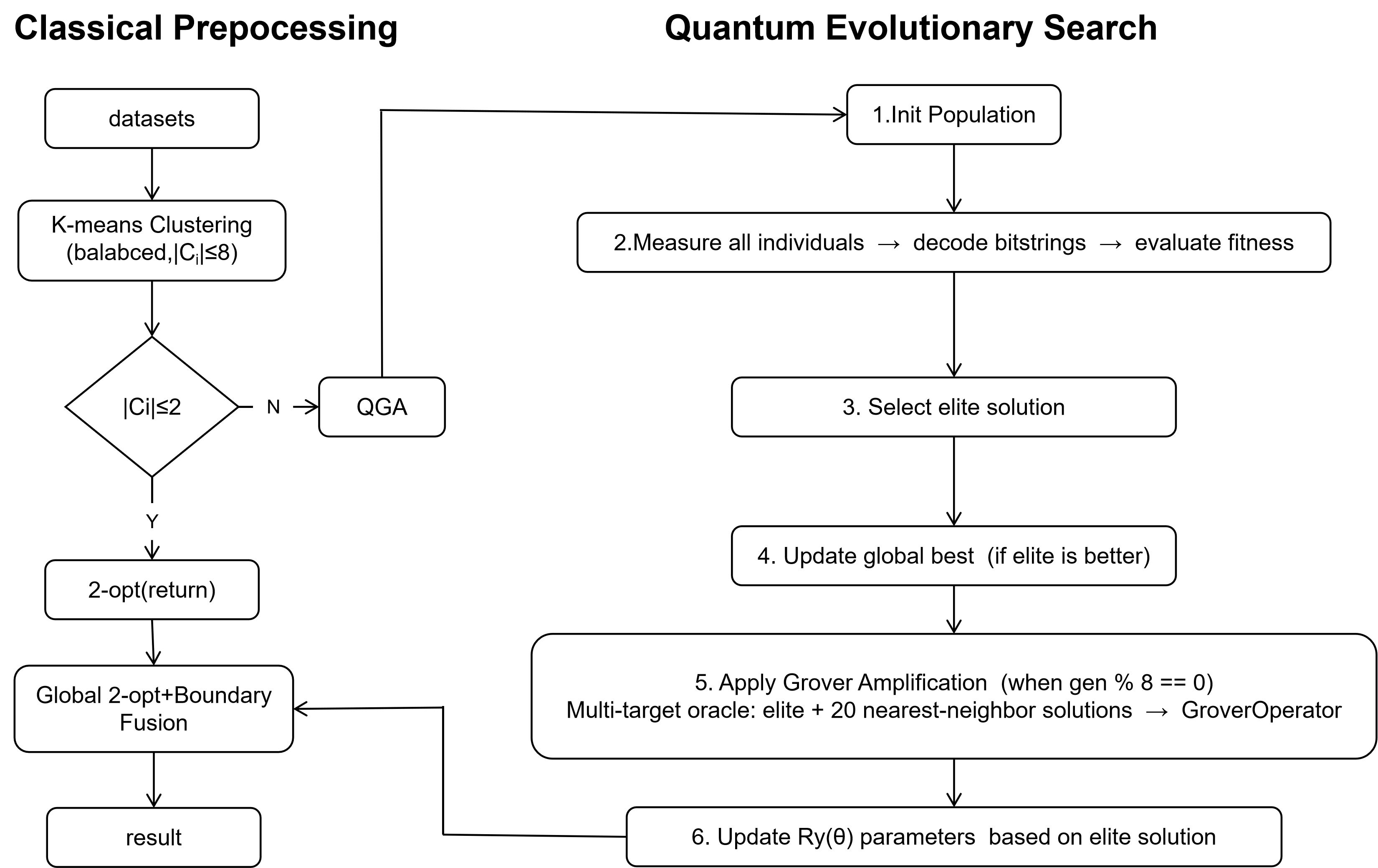}
    \caption{Hybrid Quantum-Classical Workflow.}
    \label{fig:hybrid_flowchart}
\end{figure*}

The proposed framework follows a three-stage hierarchical optimization pipeline, integrating structural decomposition, distribution-based search, and classical refinement into a unified process. An overview of the framework architecture is illustrated in Fig. \ref{fig:hybrid_flowchart}.Classicalpreprocessing handles K-means clustering and global refinement, whilequantum evolutionary search optimizes each subcluster in parallel withperiodic Grover amplification. 

Stage 1: Problem Decomposition via K-means Clustering.
In the first stage, the original problem instance is partitioned into spatially localized subproblems using clustering techniques (e.g., K-means)\cite{lu2016applying}. The decomposition proceeds until each subproblem satisfies:
\begin{equation}
|C_i| \leq M,
\end{equation}
where $M$ is a predefined threshold determined by available quantum resources, such as the number of qubits and allowable circuit depth.
    
This stage serves two key purpose:
\begin{enumerate}
\item Complexity reduction: transforming a global combinatorial problem into smaller subproblems with reduced search spaces\cite{angone2023hybrid}.
\item Hardware compatibility: explicitly bounding subproblem size to ensure feasibility under NISQ constraints.
\end{enumerate}

As a result, the original large-scale problem is mapped into a collection of subproblems, each suitable for localized optimization.

Stage 2: Quantum-Enhanced Local Optimization.
In the second stage, each subproblem is optimized independently using a hybrid quantum-classical strategy. For subproblems with $|C_i| \leq 2$, classical methods are directly applied. For subproblems with $|C_i| \geq 3$, a Grover-amplified Hybrid Quantum Genetic Algorithm (HybridQuantumGA) is employed as the primary optimization engine\cite{morimoto2024continuous}.

The optimization process within each subproblem is governed by the interaction of two mechanisms:
\begin{enumerate}
\item QGA-based probabilistic exploration, which maintains a distribution over candidate solutions and enables diverse search.
\item Amplitude amplification-based biasing, which selectively increases the sampling probability of high-quality solutions.
\end{enumerate}

Measurement and fitness evaluation provide classical feedback to iteratively update the quantum state, forming a localized probabilistic search process.

Stage 3: Global Reconstruction and Refinement.
In the final stage, locally optimized subsolutions are integrated into a global TSP route. A meta-level TSP is first constructed based on inter-cluster relationships (e.g., distances between cluster centroids), and an ordering of subproblems is determined. The local routes are then concatenated accordingly to form an initial global solution.To address discontinuities introduced by decomposition, a cross-subproblem boundary fusion mechanism is applied within a local window to refine connections between adjacent subpaths. Finally, a global 2-opt refinement is performed to eliminate crossing edges and improve overall solution quality\cite{uddin2023improvement} . This stage ensures that locally optimized structures are transformed into a globally coherent and high-quality solution.

\subsection{\textbf{\textbf{ Method Formulation and Key Mechanisms}}}
This section formalizes the internal mechanisms of the proposed framework. Rather than treating the optimization process as a sequence of isolated operations, the proposed framework interprets is formulated as a distribution-driven optimization process, where a probability distribution over candidate solutions is iteratively updated through the interaction between quantum state evolution and classical feedback.

\subsubsection{Solution Encoding and Representation }
For a subproblem containing m cities, candidate solutions are encoded using a binary representation scheme\cite{lucas2014ising}. Each city is assigned a unique identifier, and a feasible route is represented as a sequence of such identifiers\cite{papalitsas2019qubo}.  The quantum register consists of n qubits, where n is determined by the encoding length. The quantum state is represented as:
\begin{equation} \label{eq:psi}
    |\psi\rangle = \sum_{x \in \mathcal{S}} c_x |x\rangle
\end{equation}
where $\mathcal{S}$ denotes the set of encoded candidate solutions and
$c_x$ represents the probability amplitude associated with solution $x$.

This formulation defines a compact probabilistic representation of the
solution space, where the amplitudes implicitly encode a distribution over
candidate solutions.Unlike explicit population-based methods, the search is conducted over this distribution through controlled state transformations. To ensure valid decoding, infeasible encodings are mapped to valid city indices using a correction mechanism, maintaining robustness of the representation. 

\subsubsection{Quantum Genetic Update Mechanism }
The evolution of the solution distribution is governed by parameterized quantum transformations. At iteration t, the quantum state is updated as:
\begin{equation} \label{eq:state_update}
    \left|\psi_{t+1}\right\rangle = U(\theta_t) \left|\psi_t\right\rangle
\end{equation}

where $U(\boldsymbol{\theta}_t)$ denotes a parameterized unitary operator composed of
rotation gates, and $\boldsymbol{\theta}_t$ represents the tunable parameters.

The update mechanism proceeds as follows:
\begin{enumerate}
\item The quantum state is measured to generate candidate solutions.
\item Each solution is evaluated via the fitness function $f(\mathbf{x})$.
\item High-quality solutions are identified and used to adjust $\theta_t$.
\item The transformation $U(\theta_t)$ reshapes the probability distribution.
\end{enumerate}

This process can be interpreted as a distribution-level learning mechanism, where the optimization operates on the probability amplitudes rather than individual solutions. Compared to classical genetic operators, this formulation enables a more compact and globally coupled update of the search distribution, improving exploration efficiency while maintaining diversity\cite{cerezo2021variational}.  

\subsubsection{Grover-based Amplitude Amplification }
To accelerate convergence, the framework incorporates amplitude amplification as a structured enhancement of the distribution update process. Importantly, amplitude amplification is not applied as a full global search, but as a localized and periodic operator embedded within the QGA loop.

Let $M \subset S$ denote the set of high-quality candidate
solutions identified during evaluation. The corresponding oracle operator $O$
marks these states by phase inversion:
\begin{equation}
    O\lvert x\rangle =
    \begin{cases}
        -\lvert x\rangle, & x \in M, \\
         \lvert x\rangle,  & x \notin M.
    \end{cases}
\end{equation}
The diffusion operator $D$ performs inversion about the mean amplitude. The
combined operator is:
\begin{align}
    G \equiv D \cdot O
\end{align}
The diffusion operator $D$ performs inversion about the mean amplitude, and the combined operator is $G \equiv D \cdot O$~\cite{nielsen2010quantum}.
Specifically:
\begin{enumerate}
    \item It operates on subproblems of bounded size.
    \item Only a limited number of amplification steps is applied.
    \item It is invoked every $\Delta G$ iterations.
\end{enumerate}

\begin{figure}
    \centering
    \includegraphics[width=1\linewidth]{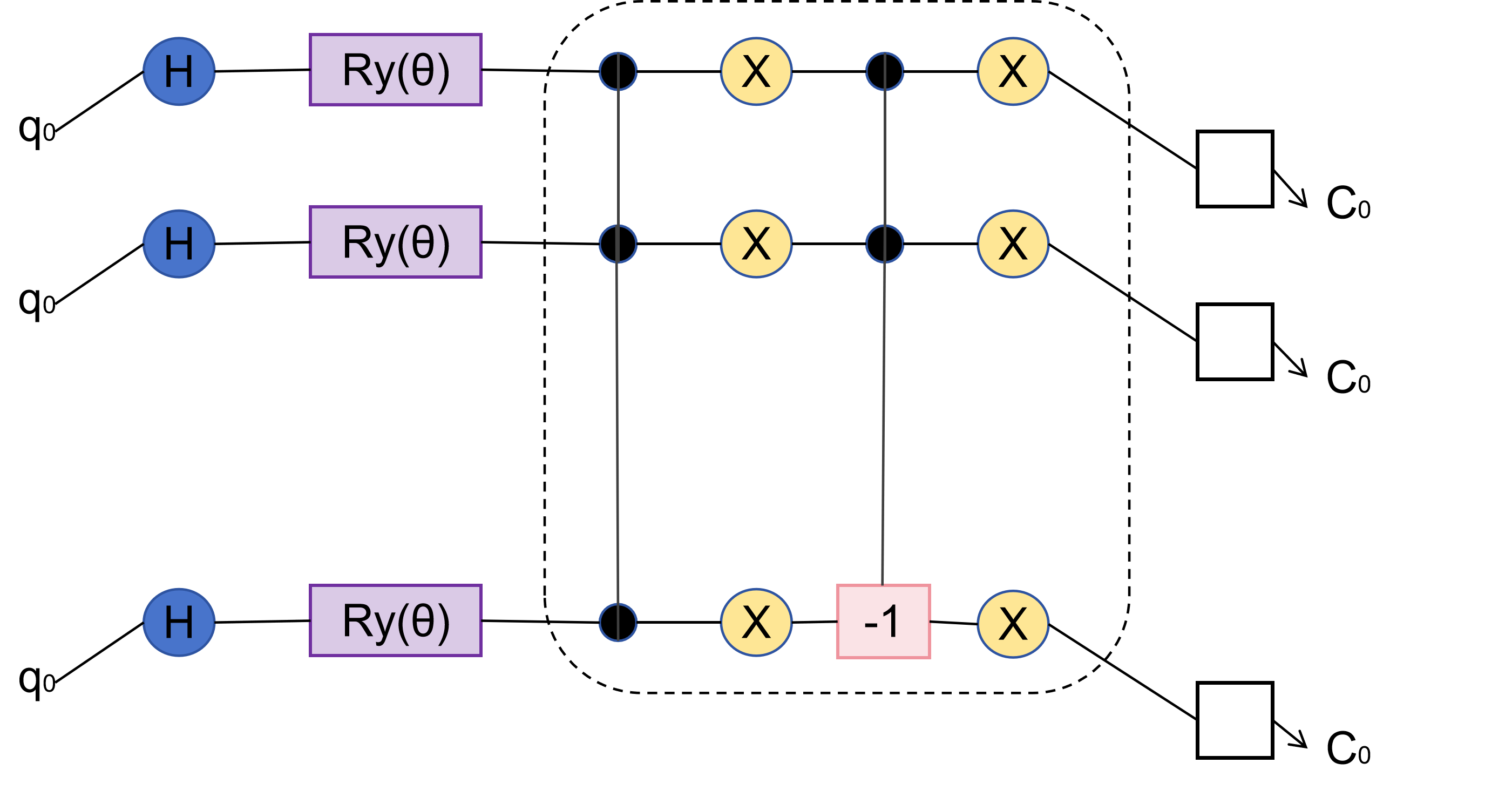}
    \caption{
Localized Grover Amplitude Amplification Circuit}
    \label{fig:2}
\end{figure}

The oracle is constructed to mark multiple high-quality states simultaneously.
Instead of relying on a single elite solution, the framework generates a
neighborhood of $K = 20$ perturbations (e.g., swap and 2-opt variants) and
marks all corresponding states. This multi-target design broadens the
attraction region of amplification, preventing premature concentration and
stabilizing convergence\cite{boyer1998tight}. Furthermore, amplitude amplification is only
activated when the subproblem size satisfies $n \leq 30$ qubits, ensuring that
circuit depth remains compatible with NISQ constraints. An illustration of the Grover circuit is provided in Fig. \ref{fig:2}.

From a functional perspective, amplitude amplification acts as a controlled
redistribution operator over the solution distribution, increasing probability
mass in promising regions while preserving exploration. As demonstrated in
Section IV, the parameter $\Delta G$ governs the trade-off between
exploration and exploitation. In this work, $\Delta G$ is empirically set to
$8$.

\subsubsection{Fitness Evaluation and Sampling }
At each iteration, the quantum state is measured to generate classical candidate solutions:
\begin{equation}
    x \sim P(x) = \lvert c_x \rvert^{2}.
\end{equation}
The fitness function is defined as:
\begin{equation}
    f(x) = \frac{1}{d(x)}.
\end{equation}
where $d(x)$ is the total path length of solution $x$.

Due to the probabilistic nature of measurement, multiple samples are drawn at each iteration. The best-performing solutions are retained to provide stable feedback for updating the quantum state. This sampling–evaluation process constitutes the classical feedback channel that drives the evolution of the quantum distribution.

\subsubsection{Hybrid Optimization Loop }
The overall optimization process forms a closed-loop hybrid system, in which quantum and classical components interact iteratively. At each iteration, the following steps are performed:
\begin{enumerate}
    \item \textbf{Sampling:} Measure the quantum state to generate candidate solutions.
    \item \textbf{Evaluation:} Compute fitness values using classical computation.
    \item \textbf{Selection:} Identify high-quality solutions.
    \item \textbf{Update:} Apply quantum transformations (QGA + amplitude amplification).
    \item \textbf{Iteration:} Repeat until convergence.
\end{enumerate}

In this formulation, quantum components govern probabilistic exploration and distribution shaping,
while classical components ensure accurate evaluation and structural refinement.
From a unified perspective, the proposed framework can be interpreted as a hybrid distribution
optimization system, where:
\begin{enumerate}
    \item QGA defines the evolution dynamics of the distribution.
    \item Amplitude amplification introduces structured bias toward high-quality regions.
    \item Classical evaluation and refinement enforce feasibility and global consistency.
\end{enumerate}

This integrated mechanism enables efficient optimization within bounded subspaces, achieving
scalability while remaining compatible with NISQ hardware constraints.

\subsection{\textbf{\textbf{Resource-Bounded Scalability and Innovation}}}
Rather than evaluating individual components in isolation, the analysis presented in this section focuses on how their interaction defines the overall efficiency and behavior of the system, and highlights the key mechanisms that enable scalable hybrid optimization under NISQ constraints.

\textbf{Resource-bounded scalability.} A defining property of the proposed framework is that its computational scaling is governed by subproblem size rather than global problem size. By enforcing an upper bound on cluster size through decomposition, the quantum resource requirement (e.g., number of qubits and circuit depth) remains constant as the problem grows. Scalability is therefore achieved by increasing the number of subproblems, rather than expanding the complexity of individual quantum computations. This design effectively transforms the global combinatorial complexity into a structured composition of bounded subproblems, where each subproblem is of fixed complexity O($m$!) with $m \ll n$. Importantly, this is not merely a divide-and-conquer heuristic, but a resource-constrained reformulation of the optimization problem, where computational feasibility is explicitly aligned with hardware limitations. As a result, the framework achieves a form of resource-invariant scaling, which is particularly suitable for NISQ environments    \cite{mcclean2016theory}.

\textbf{Controlled quantum enhancement.}Amplitude amplification is incorporated as a localized and periodic operator, rather than a global search mechanism. This design avoids deep quantum circuits and eliminates the need for constructing large-scale oracles. By selectively increasing the probability of high-quality solutions, amplitude amplification improves convergence behavior while preserving exploration\cite{lamza2024qhyper}. The periodic embedding strategy ensures a balance between enhancement and diversity, as further validated in the ablation study (Section IV).

\textbf{Closed-loop hybrid optimization.} The proposed framework operates as a tightly
coupled quantum--classical system, where optimization emerges from iterative interaction
rather than isolated computation. Quantum components maintain and evolve a probabilistic
representation of candidate solutions, while classical components provide deterministic
evaluation and structural refinement. This interaction forms a feedback-driven
optimization loop: sampling $\rightarrow$ evaluation $\rightarrow$ distribution update
$\rightarrow$ refinement. Unlike purely classical heuristics or standalone quantum
algorithms, performance gains arise from this closed-loop coupling. Quantum mechanisms
enhance exploration and guide the search distribution, while classical processes enforce
feasibility and improve structural coherence. The resulting system can be viewed as a
hybrid distribution optimization framework, in which global solution quality emerges from
the coordinated dynamics of probabilistic search and deterministic refinement.

\section{Experimental Evaluation}
To systematically validate the proposed NISQ-aware hybrid optimization framework, we conduct a comprehensive experimental study evaluating solution quality, scalability, and robustness under realistic constraints.The experimental design is explicitly aligned with the key mechanisms introduced in Section III, enabling direct validation of decomposition-based scalability, quantum-enhanced local search, and hybrid optimization dynamics.
\subsection{Experimental Setup}
Datasets. To comprehensively evaluate the proposed framework, experiments are conducted on two types of datasets:

Benchmark datasets. We select 10 standard instances from the TSPLIB benchmark datasets, covering a range of problem scales:
\begin{enumerate}
    \item \textit{Small-scale:} Ulysses-16, Ulysses-22
    \item \textit{Medium-scale:} Bayg-29, Bays-29, Dantzig-42, Att-48, Eil-51, Berlin-52
    \item \textit{Large-scale:} Eil-76, St-70
\end{enumerate}

These datasets exhibit diverse spatial distributions, enabling evaluation under different structural conditions. For each instance, all algorithms are executed independently 10 times. The average path length, best solution, and standard deviation are recorded. Known optimal solutions provided by TSPLIB are used as reference benchmarks.

Synthetic datasets. To systematically evaluate scalability, we construct synthetic datasets with problemsizes ranging from ten to 100 cities. City coordinates are generated from a uniformdistribution within a two-dimensional plane. For each problem size, ten independentinstances are generated, and each instance is evaluated over ten runs. Reportedresults correspond to averaged values, reducing stochastic variability and ensuringstatistical robustness. 

Baseline Methods. We compare the proposed framework with representative classical and quantum-inspired baselines:
\begin{enumerate}
\item Ant Colony Optimization (ACO), using standard pheromone update rules
\item Genetic Algorithm (GA), with tournament selection, partially-mapped crossover (PMX), and mutation
\item Greedy algorithm, based on nearest-neighbor selection
\item Quantum Ant Colony Optimization (QACO), incorporating clustering-based decomposition.
\end{enumerate}

All baseline methods are implemented using commonly adopted configurations, with parameters tuned to ensure competitive performance.

Implementation Details. All experiments are conducted using Qiskit-based quantum simulation on classical computing hardware. Euclidean distance is used as the cost metric for all TSP instances.

For the proposed framework, key parameters are configured as follows:
\begin{enumerate}
\item The subproblem size threshold $M$ is set to $8$, constrained by qubit limits.
\item The Grover-enhanced QGA runs for $28$ generations per subproblem, with $2$-opt disabled during evolution and applied only in the final refinement stage.
\item Amplitude amplification is applied periodically with interval $\Delta G = 8$, unless otherwise specified.
\item Classical global refinement employs $2$-opt with boundary fusion (window size $= 3$--$4$).
\end{enumerate}

\textbf{Evaluation Protocol.}To ensure statistical reliability, each experiment is repeated over multiple independent runs. Performance is evaluated using both average path length and variability metrics (e.g., coefficient of variation), capturing both solution quality and stability. This protocol explicitly accounts for the stochastic nature of the hybrid optimization process.

\textbf{Design Rationale.}The experimental configuration is designed to directly validate the key mechanisms of the proposed framework across three complementary dimensions: solution quality, scalability, and robustness. The TSPLIB benchmark datasets enable evaluation of overall solution quality and provide a standard basis for comparison with established methods, while the synthetic datasets offer a controlled environment for systematically analyzing scalability under increasing problem sizes. The inclusion of both classical and quantum-inspired baselines ensures that performance differences can be attributed to the proposed hybrid optimization strategy rather than to specific algorithmic choices. In addition, multiple independent runs and variability metrics are incorporated to capture both average performance and stability, accounting for the stochastic nature of the hybrid optimization process. Overall, this experimental design ensures that the observed performance gains can be systematically linked to the core components of the framework—namely, decomposition, quantum-enhanced local search, and hybrid refinement—thereby providing a rigorous and interpretable validation of the proposed approach.

\subsection{ Overall Performance on Benchmark Datasets}
\begin{figure*}[t]
    \centering
    \includegraphics[width=\textwidth]{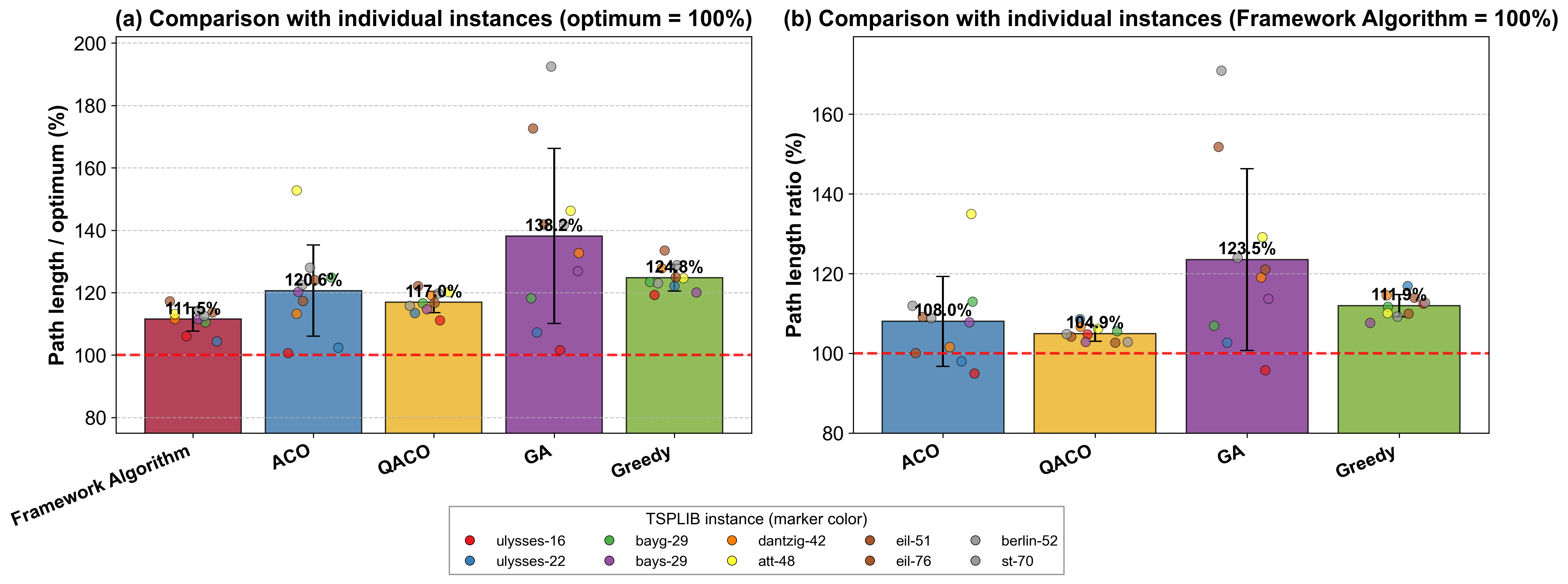}
  \caption{Comparison of route-length statistics and normalized ratios for ACO, QACO, GA, Greedy, and Hybrid Framework Algorithm.}
    \label{fig:3}
\end{figure*}

Table \ref{tab:tsplib} presents the quantitative performance comparison of all algorithms on the TSPLIB benchmark datasets, where each value corresponds to the average path length over 10 independent runs. The known optimal solutions are provided as references.Overall, the proposed framework consistently achieves the best or near-best performance across most benchmark instances, with advantages becoming increasingly pronounced as problem size grows. This trend is further supported by the statistical comparison shown in Fig. \ref{fig:3}, which visualizes both performance ratios and variance across datasets.
\begin{table*}[!htbp]
\caption{Comparative path lengths on TSPLIB datasets (average of 10 independent verification runs).}
\label{tab:tsplib}
\centering
{\small
\begin{tabular}{lccccccc}
\toprule
Dataset     & Optimum & ACO    & GA      & Greedy   & QACO     & \textbf{Framework Algorithm} \\
\midrule
Ulysses-16  & 74.11   & \textbf{74.62}  & 75.27   & 88.402   & 82.34    & 78.57\\
Ulysses-22  & 75.67   & \textbf{77.48}  & 81.19   & 92.404   & 85.84    & 79.05 \\
Bayg-29     & 9074.15 & 11334.54 & 10728.40 & 11207.54 & 10587.25 & \textbf{10029.16} \\
Bays-29     & 9291.35 & 11173.94 & 11792.14 & 11163.612 & 10662.26 & \textbf{10367.53} \\
Dantzig-42  & 699.00  & 791.91 & 928.13  & 893.554  & 832.07   & \textbf{779.56} \\
Att-48      & 33523.71 & 51225.25 & 49025.50 & 41789.203 & 40296.72 & \textbf{37939.95} \\
Eil-51      & 429.98  & 504.43 & 610.18  & 574.345  & 525.32   & \textbf{504.03} \\
Eil-76      & 545.39  & 676.85 & 941.49  & 681.805  & 636.62   & \textbf{620.02} \\
Berlin-52   & 7544.37 & 9657.42 & 10693.30 & 9721.307 & 9039.85 & \textbf{8623.663} \\
St-70       & 678.60  & 831.39 & 1306.07 & 834.722  & 786.13   & \textbf{764.049} \\
\bottomrule
\end{tabular}
}
\end{table*}
\textbf{Performance across different problem scales.}A clear scale-dependent pattern is observed. For small-scale instances such as Ulysses-16 and Ulysses-22, classical methods (e.g., ACO) achieve results closer to the optimal solution, as the search space remains limited and can be effectively explored by heuristic strategies.

As the problem size increases, the advantage of the proposed framework becomes increasingly pronounced. For medium- and large-scale instances (e.g., Bayg-29, Bays-29, Att-48, Berlin-52, Eil-76, and St-70), the proposed framework consistently outperforms most baselines and frequently achieves the best results. For example, on the St-70 dataset, the Grover-enhanced QGA achieves a path length of 764.05, compared to 831.39 for ACO, 1306.07 for GA, and 786.13 for QACO. This trend is also reflected in Fig. 4(a), where the performance ratio relative to the optimal solution remains consistently lower for the proposed framework as the problem size increases. This scale-dependent behavior directly reflects the design principle of the framework, where performance gains emerge more prominently as the search space becomes sufficiently large.

\textbf{Relative performance and stability.}Fig. \ref{fig:3}(b) provides a relative comparison using the proposed framework as the reference baseline. All competing algorithms show higher relative path lengths on average across datasets, with larger deviations observed for GA and Greedy methods.

In addition to improved average performance, the proposed framework demonstrates lower variance across independent runs, as indicated by the error bars in Fig. \ref{fig:3}. This suggests that the framework not only achieves better solutions but also maintains stable performance under stochastic optimization.

\begin{figure*}[t]
  \centering
  \includegraphics[width=\textwidth]{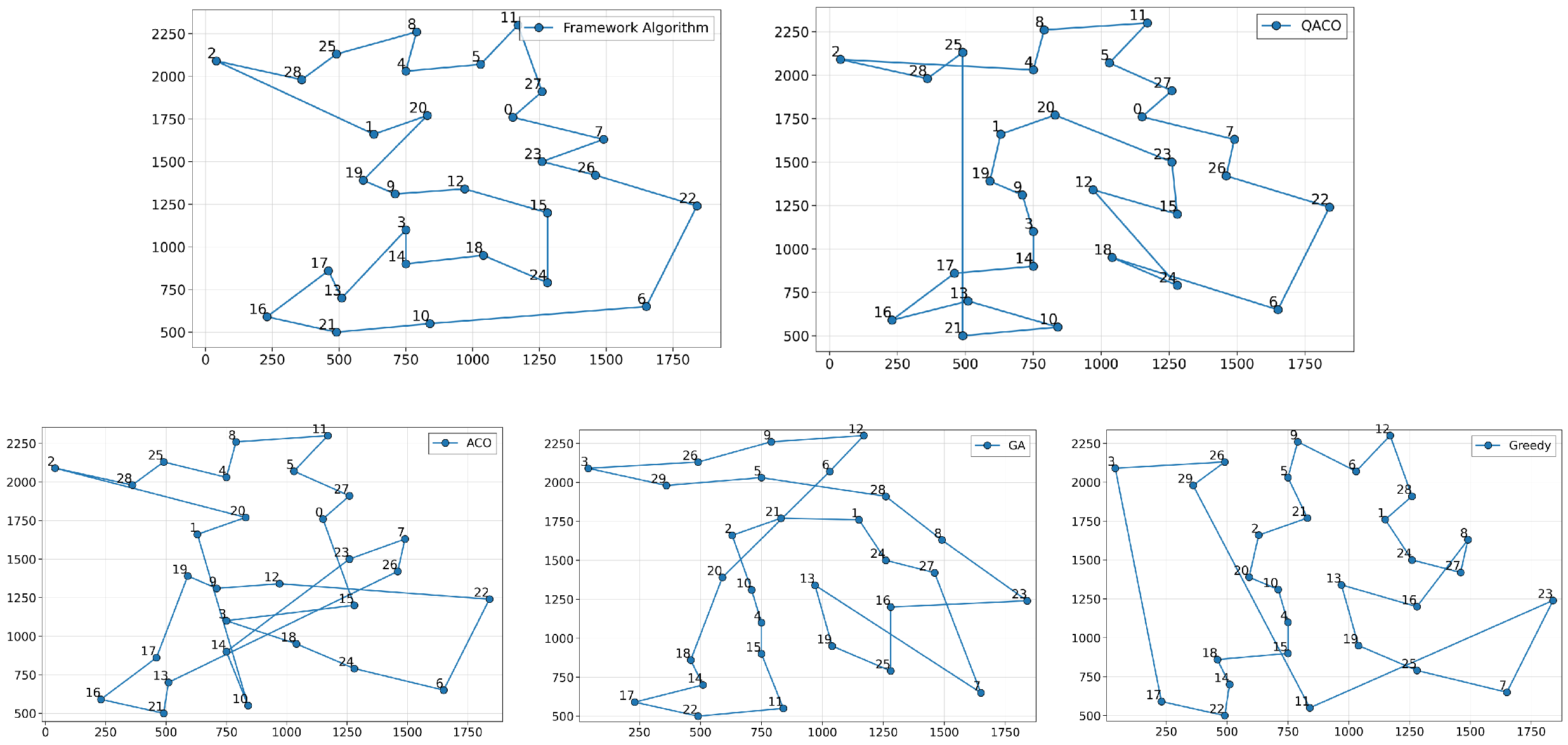}
  \caption{Route geometry comparison on the Bays-29 dataset for the proposed hybrid framework, QACO, ACO, GA, and Greedy methods. }
  \label{fig:placeholder}
\end{figure*}

\textbf{Structural quality of Geometric properties of solutions.}Beyond numerical results, the geometric properties of the obtained routes are illustrated in Fig. \ref{fig:placeholder} using the Bays29 dataset. The paths generated by the proposed framework exhibit smoother transitions and fewer crossings, indicating improved global consistency. In contrast, classical methods such as GA and Greedy produce more irregular routes with visible intersections, whereas reflecting their limitations in maintaining global coherence.





\textbf{Mechanism-level interpretation.}The observed performance improvements are consistent with the design of the proposed framework.
\begin{enumerate}
    \item \textbf{Decomposition reduces effective search complexity.} The clustering-based
          divide-and-conquer strategy partitions the original problem into smaller subproblems,
          reducing the search space from $O(n!)$ to a set of tractable local problems.

    \item \textbf{Quantum-enhanced search improves convergence efficiency.} Within each subproblem,
          the Grover-enhanced QGA biases the sampling distribution toward high-quality solutions,
          accelerating convergence compared to classical evolutionary approaches.

    \item \textbf{Hybrid optimization ensures global consistency.} The combination of
          quantum-enhanced local search and classical global refinement enables a balance
          between exploration and exploitation, resulting in improved solution quality and stability.
\end{enumerate}

\textbf{Comparison with decomposition-based baselines.}A key comparison is with QACO, which also employs clustering-based decomposition. Despite sharing a similar divide-and-conquer strategy, the proposed framework consistently achieves better performance across most datasets. This indicates that decomposition alone is insufficient, and that performance gains arise from the integration of quantum evolutionary mechanisms and amplitude amplification within the optimization process.

\subsection{Scalability on Synthetic Datasets}
\begin{figure}[!htbp] 
    \centering
    \includegraphics[width=\columnwidth]{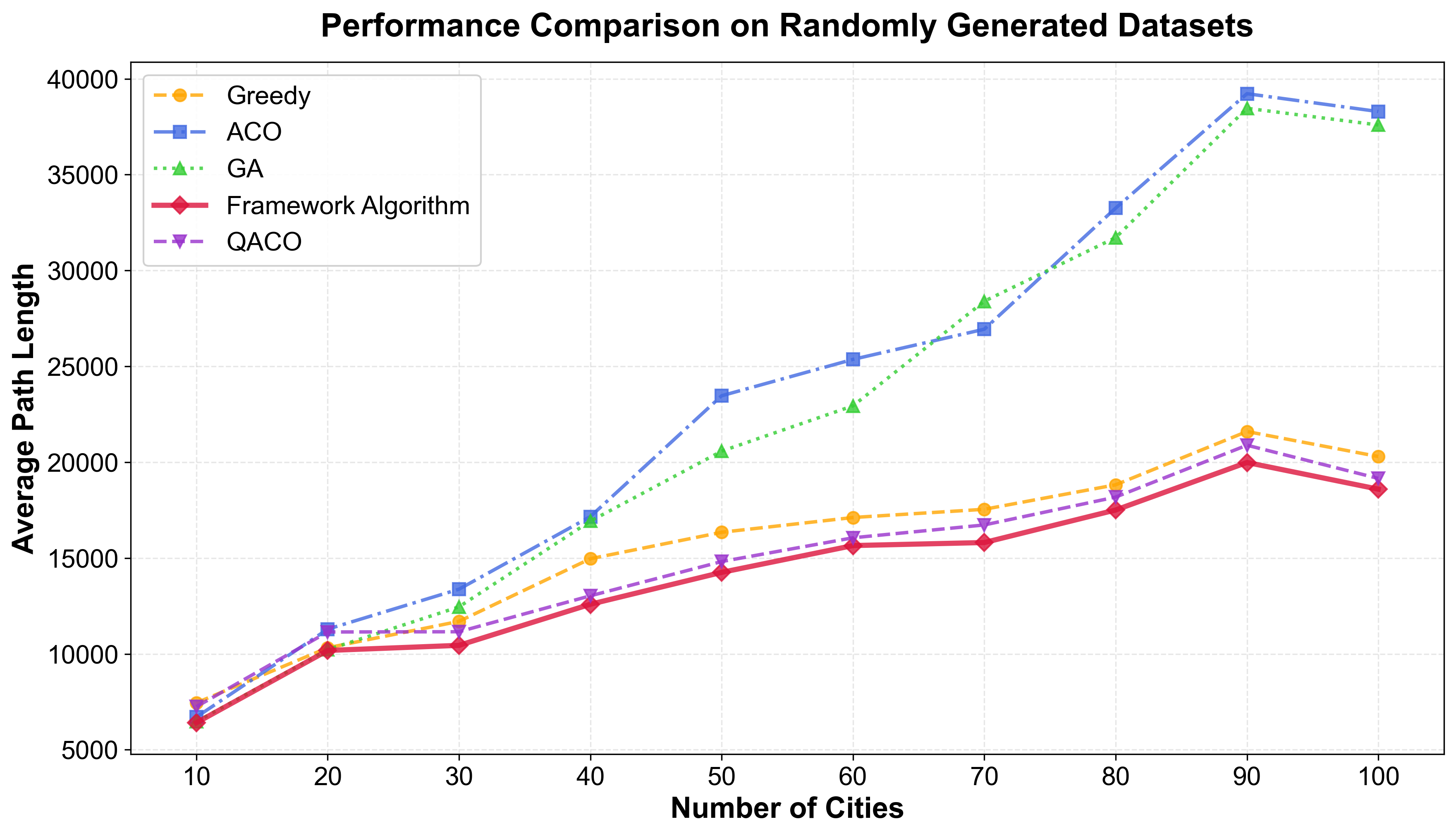}
  \caption{Algorithm comparison on random datasets.}
    \label{fig:random_dataset_comparison}
\end{figure}

To systematically evaluate scalability, we conduct experiments on randomly generated datasets with problem sizes ranging from 10 to 100 cities. The quantitative results are summarized in Table \ref{tab:algorithm_comparison}, and the overall scaling trends are visualized in Fig. \ref{fig:random_dataset_comparison}.
\begin{table*}[!htbp]
\centering
\caption{Average path lengths for each algorithm on randomly generated datasets
(number of cities from 10 to 100). Values show mean $\pm$ CV(\%);
bold text indicates the best mean result at this scale.}
\label{tab:algorithm_comparison}
{\small
\begin{tabular}{lcccccc}
\toprule
Dataset & Greedy & ACO & GA & QACO & \textbf{Framework Algorithm} \\ \midrule
10  & 7446.4$\pm$4.0   & 6701.5$\pm$5.6   & 6463.2$\pm$7.7   & 7244.0$\pm$6.2   & \textbf{6407.1$\pm$2.7}   \\
20  & 10319.4$\pm$3.3  & 11289.8$\pm$6.2  & 10220.8$\pm$5.1  & 11142.6$\pm$6.0  & \textbf{10178.7$\pm$5.5}  \\
30  & 11696.6$\pm$15.4 & 13375.3$\pm$7.0  & 12454.1$\pm$6.6  & 11155.0$\pm$5.7  & \textbf{10445.0$\pm$4.3}  \\
40  & 14964.1$\pm$8.5   & 17161.3$\pm$7.4  & 16924.4$\pm$5.8  & 13028.8$\pm$3.5  & \textbf{12588.6$\pm$4.5}  \\
50  & 16344.8$\pm$5.0   & 23466.8$\pm$4.2  & 20576.4$\pm$11.2 & 14817.0$\pm$5.6  & \textbf{14252.8$\pm$3.0}  \\
60  & 17114.9$\pm$5.9   & 25363.5$\pm$6.7  & 22940.2$\pm$6.8  & 16056.9$\pm$5.0  & \textbf{15651.2$\pm$4.9}  \\
70  & 17543.3$\pm$6.0   & 26939.0$\pm$5.2  & 28398.7$\pm$11.1 & 16726.9$\pm$4.8  & \textbf{15808.8$\pm$2.5}  \\
80  & 18820.8$\pm$3.6   & 33256.1$\pm$4.7  & 31717.1$\pm$6.1  & 18179.5$\pm$5.2  & \textbf{17507.0$\pm$4.1}  \\
90  & 21600.8$\pm$2.2   & 39225.3$\pm$4.0  & 38463.8$\pm$6.3  & 20886.2$\pm$3.3  & \textbf{19985.3$\pm$3.6}  \\
100 & 20295.3$\pm$5.1   & 38293.2$\pm$6.5  & 37592.2$\pm$3.1  & 19153.4$\pm$3.5  & \textbf{18593.2$\pm$2.9}  \\ \bottomrule
\end{tabular}
}
\end{table*}

\textbf{Scaling behavior across increasing problem sizes.}A clear scale-dependent performance pattern is observed. For small-scale instances (10--20 cities), all algorithms achieve comparable performance, with only marginal differences. This is expected, as the search space remains limited and can be effectively explored by classical heuristics.

As the problem size increases, the performance gap becomes increasingly pronounced. Starting from around 30--40 cities, the proposed framework consistently achieves the lowest path length across all scales. For example, at 70 cities, the proposed framework achieves a path length of 15808.8, compared to 26939.0 for ACO and 28398.7 for GA. This advantage persists up to 100 cities, where the proposed framework continues to outperform all baselines.

This trend is illustrated in Fig. 5, where the curve corresponding to the proposed framework exhibits a consistently lower growth rate compared to other algorithms.

\textbf{Variance and stability analysis.}Table \ref{tab:algorithm_comparison} also reports the coefficient of variation (CV), providing insight into algorithm stability. The proposed framework consistently achieves lower CV values across all scales (e.g., 2.5\% at 70 cities), compared to higher variability in GA and ACO. This indicates that the proposed framework not only improves average solution quality but also maintains stable performance under stochastic optimization, which is particularly important for large-scale problems.

\textbf{Mechanism-level explanation of scalability.}The observed scalability stems from the structural design of the proposed framework.Specifically, the framework effectively decouples global problem complexity from local optimization complexity. This decoupling is realized through the following mechanisms: 
\begin{enumerate}
\item \textbf{Decomposition enables complexity control.} The K-means-based divide-and-conquer strategy partitions the original problem into subproblems of bounded size, effectively transforming the global search space from $O(n!)$ to a set of smaller problems with complexity $O(m!)$, where $m \ll n$. This significantly limits the growth of computational complexity.

\item \textbf{Localized optimization improves efficiency.} Independent optimization within each subproblem avoids global combinatorial explosion and enables implicit parallelism, further improving scalability.

\item \textbf{Quantum-enhanced search accelerates convergence.} The Grover-enhanced QGA biases the sampling distribution toward high-quality solutions within each subproblem, reducing the number of iterations required f or convergence.
\end{enumerate}

\textbf{Comparison with baseline methods.}The scalability advantage is further clarified through comparison with baseline algorithms. Classical methods such as ACO and GA exhibit rapid performance degradation as the problem size increases, reflecting their limitations in handling exponentially growing search spaces.

 In contrast, both the proposed framework and QACO demonstrate more stable scaling behavior due to decomposition. However, the proposed framework consistently outperforms QACO across all scales, indicating that decomposition alone is insufficient, and that the integration of quantum-enhanced evolutionary search and amplitude amplification plays a critical role.

\subsection{ Robustness under Noise}
To evaluate the practical feasibility of the proposed framework under realistic quantum conditions, we simulate noise using Qiskit, incorporating both bit-flip noise and thermal relaxation noise. Noise levels are varied from 2\% to 10\% to reflect typical NISQ environments.

The experimental results are summarized in Table \ref{NoiseTab1}. Across all datasets, the degradation in solution quality remains limited, and the obtained path lengths stay close to those of noise-free simulations. For example, on the Bayg-29 dataset, the path length varies within a relatively narrow range across different noise levels, indicating stable performance under noise perturbations.
\begin{table*}[!htbp]
\centering
\caption{The results of datasets with noises.}
\label{NoiseTab1}
{\small
\begin{tabular}{lcccccc} \cmidrule{1-7}
Dataset& No noise & 2\%  & 4\%   & 6\%   & 8\%   & 10\%  \\ \midrule
Ulysses-22 & 79.05    & 80.39 & 80.23 & 80.50 & 81.01 & 78.20  \\
Bayg-29    & 10029.16 & 9641.24 & 9950.68 & 9879.24 & 10032.89 & 10283.70 \\
Eil-51     & 504.03   & 532.75 & 519.02 & 504.44 & 503.24 & 496.78  \\
Eil-76     & 620.02   & 632.70 & 616.07 & 613.67 & 623.73 & 639.76  \\
KroA100    & 24208.68 & 25047.59 & 24125.59 & 23029.09 & 24690.64 & 25149.97 \\ \bottomrule
\end{tabular}
}
\end{table*}
\textbf{Noise robustness characteristics.}A key observation is that the deviation from noise-free results does not increase monotonically with noise intensity. This suggests that noise-induced errors do not accumulate significantly during the optimization process.

This behavior can be attributed to three structural properties of the proposed framework:
\begin{enumerate}
\item Shallow circuit design. The framework operates on decomposed subproblems with limited qubit counts, resulting in relatively shallow quantum circuits and reduced error accumulation.
\item Localized quantum operations. Quantum processing is restricted to subproblems, preventing error propagation across the global solution space.
\item Measurement-driven feedback. The hybrid optimization loop relies on repeated measurement and classical evaluation, which effectively filters out low-quality solutions and mitigates the impact of noise-induced perturbations.
\end{enumerate}

\textbf{Stability under different noise models.}Consistent behavior is observed under both bit-flip and thermal relaxation noise models. Even when random noise is introduced at varying levels up to 10\%, the framework maintains stable performance across datasets. This indicates that the observed robustness is not tied to a specific noise model, but is instead attributable to the underlying hybrid optimization structur

\textbf{Implications for NISQ feasibility.}The experimental findings demonstrate that the proposed framework is compatible with realistic quantum hardware conditions. By combining decomposition, shallow circuits, and controlled quantum enhancement, the framework effectively mitigates noise effects without requiring explicit error correction. Notably, robustness in this framework is not achieved by eliminating noise, but by structurally limiting its impact on the optimization process. This aligns with the design philosophy of adapting algorithms to hardware constraints, rather than relying on fault-tolerant quantum computation.

\subsection{ Ablation Study and Mechanism Validation}

To quantify the contribution of individual components and validate the mechanism-level design introduced in Section III, we conduct ablation experiments by selectively removing or replacing key modules of the proposed framework. We emphasize that the clustering-based decomposition is not an optional heuristic but a hardware-driven necessity under NISQ qubit-budget constraints, as it bounds each subproblem to a feasible register size and circuit depth. Accordingly, this decomposition mechanism is regarded as a fundamental enabling component of the proposed framework and is excluded from the ablation settings.

The study considers three configurations:
\begin{enumerate}
    \item \textbf{Full Model:} the complete proposed hybrid framework.
    \item \textbf{w/o Amplification:} removing periodic Grover-based amplitude amplification.
    \item \textbf{w/o QGA (replaced with 2-opt):} replacing the quantum genetic algorithm with classical 2-opt optimization.
\end{enumerate}

\begin{figure}[H]
    \centering
    \includegraphics[width=\linewidth]{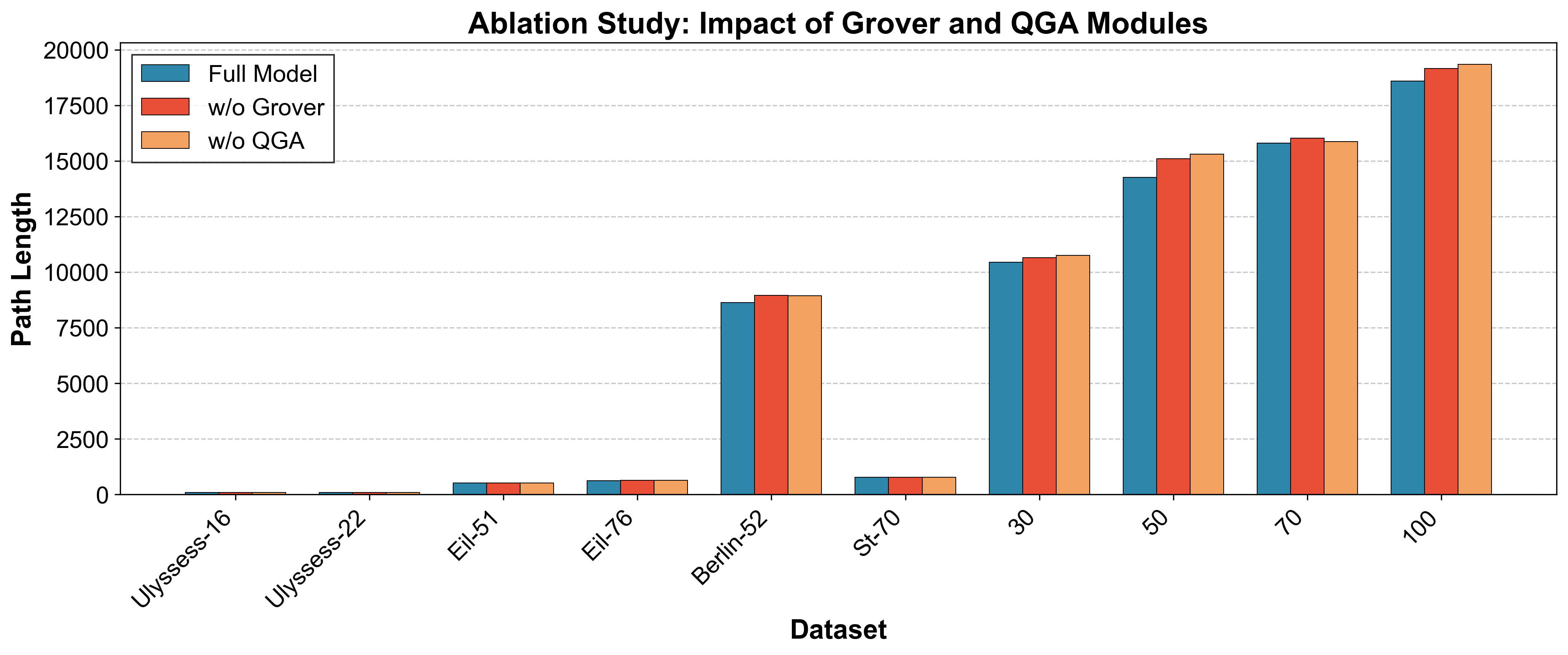}
    \caption{Ablation comparison of amplification and QGA modules.}
    \label{fig:5}
\end{figure}

The results are reported in Table \ref{tab:ablation} and visualized in Fig. \ref{fig:5}.
\begin{table*}[!htbp]
\centering
\caption{Performance Comparison of Ablation Variants}
\label{tab:ablation}
{\footnotesize
\begin{threeparttable}
\begin{tabular}{lccc}
\toprule
\textbf{Dataset} & \textbf{Full Model} & \textbf{w/o Amplification} & \textbf{w/o QGA (2-opt)} \\
\midrule
Ulysses-16   & \textbf{78.57}    & 78.84  (+0.34\%) & 78.63  (+0.08\%)  \\
Ulysses-22   & \textbf{79.05}    & 80.17  (+1.42\%) & 79.84  (+1.00\%)  \\
Eil-51       & \textbf{504.03}   & 506.52 (+0.50\%) & 507.46 (+0.68\%)  \\
Eil-76       & \textbf{620.02}   & 638.03 (+2.90\%) & 637.53 (+2.82\%)  \\
Berlin-52    & \textbf{8623.66}  & 8957.31 (+3.87\%) & 8937.16 (+3.64\%) \\
St-70        & \textbf{764.05}   & 772.86 (+1.15\%) & 776.70 (+1.66\%)  \\
Random-30    & \textbf{10444.97} & 10647.37 (+1.94\%) & 10754.08 (+2.96\%) \\
Random-50    & \textbf{14252.80} & 15102.10 (+5.96\%) & 15308.47 (+7.41\%) \\
Random-70    & \textbf{15808.77} & 16021.68 (+1.35\%) & 15868.02 (+0.37\%) \\
Random-100   & \textbf{18593.20} & 19163.62 (+3.07\%) & 19347.01 (+4.05\%) \\
\midrule
Avg. Degradation & --- & +2.25\% & +2.47\% \\
\bottomrule
\end{tabular}
\begin{tablenotes}
\item \textit{Note:} Percentage values in parentheses indicate performance degradation from the Full Model. ``w/o Amplification'' refers to the removal of the periodic amplitude amplification mechanism from the full framework.
\end{tablenotes}
\end{threeparttable}
}
\end{table*}

\textbf{Overall degradation and scale dependence.}Across all datasets, both ablated variants consistently underperform the Full Model. The average degradation is +2.25\% (without Amplification) and +2.47\% (without QGA), indicating that both components contribute non-trivially to performance.

More importantly, the degradation exhibits a clear scale-dependent pattern. For small-scale instances (e.g., Ulysses-16), the impact of removing either component is minimal. However, as the problem size increases, the performance gap becomes increasingly pronounced. For example:
\begin{enumerate}
    \item \textbf{Random-50:} +5.96\% (w/o Amplification), +7.41\% (w/o QGA).
    \item \textbf{Random-100:} +3.07\% (w/o Amplification), +4.05\% (w/o QGA).
\end{enumerate}

This trend indicates that the removed mechanisms become increasingly critical as the solution space grows, providing direct empirical evidence for their role in large-scale optimization.

\textbf{Role of amplitude amplification:}convergence control. Removing amplitude amplification leads to consistent degradation across all datasets, with a more pronounced impact on medium-to-large instances (e.g., Berlin-52: +3.87\%, Eil-76: +2.90\%).

As shown in Fig. 6, the without Amplification configuration systematically produces longer path lengths compared to the Full Model. This confirms that amplitude amplification acts as a controlled biasing mechanism, accelerating convergence by increasing the probability of high-quality candidate solutions. The effect becomes more significant for larger problem instances, where the search space is more complex and targeted biasing becomes essential for efficient convergence.

\textbf{Role of QGA: exploration capability.}Replacing QGA with classical 2-opt results in slightly larger degradation (+2.47\% on average), particularly for larger instances:
\begin{enumerate}
    \item \textbf{Random-50:} +7.41\%.
    \item \textbf{Berlin-52:} +3.64\%.
\end{enumerate}

This indicates that QGA plays a critical role in maintaining global exploration capability. While 2-opt is effective for local refinement, it lacks the ability to explore diverse regions of the solution space, leading to premature convergence and suboptimal solutions. These results demonstrate that QGA is essential for sustaining diversity in the probabilistic search process.

\textbf{Functional distinction and synergy.}A direct comparison between the two ablation variants reveals a clear functional distinction:
\begin{enumerate}
    \item Amplitude amplification improves convergence efficiency.
    \item QGA enhances exploration and diversity.
\end{enumerate}

The slightly larger degradation observed when removing QGA suggests that exploration is more critical in large-scale combinatorial optimization, whereas amplification provides complementary acceleration.

Notably, neither component alone can recover the performance of the Full Model. The consistent performance gap observed in Fig. 6 indicates that the improvements arise from the synergistic interaction between the two mechanisms. QGA defines a diverse probabilistic search distribution, while amplitude amplification selectively reinforces high-quality regions. Their interaction enables an effective balance between exploration and exploitation.

\textbf{Effect of amplification frequency ($\Delta G$). }To further analyze the role of amplitude amplification, we study the effect of the embedding frequency parameter $\Delta G$, with results summarized in Table \ref{tab:grover-frequency}.

The results reveal a clear unimodal relationship between $\Delta G$ and solution
quality: an optimal $\Delta G$ exists, with solution quality peaking at
$\Delta G = 8$ and degrading on both sides. Excessive amplification
($\Delta G = 4$) leads to reduced diversity and slight degradation, whereas
insufficient amplification ($\Delta G = 12, 16$) weakens convergence and
degrades performance. This behavior reflects the trade-off between exploration
and exploitation: excessive amplification leads to premature concentration of
probability mass, whereas insufficient amplification fails to effectively guide
convergence. The optimal setting $\Delta G = 8$ achieves a balanced interaction
between these two effects.

\textbf{Mechanism-level validation.}The ablation results provide direct empirical validation of the framework design:
\begin{enumerate}
    \item QGA enables effective exploration in high-dimensional solution spaces.
    \item Amplitude amplification improves convergence toward high-quality solutions.
    \item Their interaction forms a coherent hybrid optimization mechanism.
\end{enumerate}

The consistent degradation observed across all ablated variants confirms that these components are essential and non-redundant, and that performance gains arise from their coordinated interaction.
\begin{table}[htbp]
\centering
\caption{Impact of amplitude amplification embedding frequency $\Delta G$ on solution quality.}
\label{tab:grover-frequency}
\scriptsize

\resizebox{\columnwidth}{!}{%
\begin{tabular}{lcccc}
\toprule
Dataset & $\Delta G$=4 & $\Delta G$=8 & $\Delta G$=12 & $\Delta G$=16 \\
\midrule
Ulysses-16    & 78.71   & \textbf{78.57} & 78.77   & 78.65   \\
Ulysses-22    & 80.12   & \textbf{79.05} & 80.20   & 79.36   \\
Eil-51        & 505.80  & \textbf{504.03}& 505.40  & 504.37  \\
Berlin-52     & 8547.70 & \textbf{8623.66}& 8917.13 & 8914.13 \\
Eil-76        & 629.72  & \textbf{620.02}& 639.86  & 624.54  \\
St-70         & 772.68  & \textbf{764.05}& 769.06  & 773.48  \\
Random-30     & 10759.27& \textbf{10444.97}& 10671.98& 10502.06\\
Random-50     & 15091.39& \textbf{14252.80}& 15083.34& 15032.18\\
Random-70     & 15878.81& \textbf{15808.77}& 15918.33& 15918.93\\
Random-100    & 19218.78& \textbf{18593.20}& 19214.49& 19044.26\\
\bottomrule
\end{tabular}%
}

\end{table}
\section{ Discussion}
The experimental results in Section IV demonstrate that the proposed framework achieves consistent improvements in solution quality, scalability, and robustness under realistic conditions. In this section, we interpret these results from a unified perspective and discuss the underlying principles that enable these improvements.

Rather than viewing the framework as a combination of independent components, we interpret it as a structured hybrid optimization paradigm, in which decomposition, distribution-level exploration, and quantum enhancement are jointly coordinated. This perspective provides a general understanding of how hybrid quantum–classical systems can be designed under NISQ constraints.

In particular, the discussion focuses on three key aspects:
\begin{enumerate}
    \item how scalability emerges from resource-bounded decomposition,
    \item how localized enhancement is integrated in a controlled manner,
    \item how the hybrid optimization loop enables stable and effective search dynamics.
\end{enumerate}

\subsection{ Practical Implications under NISQ Constraints}
A central objective of this work is to bridge the gap between large-scale combinatorial optimization problems and the limitations of current NISQ devices. The proposed framework addresses this challenge through a resource-aware structural design, in which both problem representation and optimization procedures are explicitly adapted to hardware constraints.

First, scalability is achieved through structured decomposition. By partitioning the original problem into subproblems of bounded size ($m\le M$), the quantum resource requirement is explicitly controlled. Importantly, while the global problem size increases, the quantum resource requirement per subproblem remains constant.As summarized in Table \ref{tab:resource-stats}, even for large-scale instances (e.g., 100 cities), the maximum subcluster size remains at 8 cities, corresponding to at most 24 qubits per subproblem.  This reflects a resource-bounded scaling mechanism, in which scalability is achieved through increasing the number of tractable subproblems rather than expanding circuit complexity.

Second, the framework adopts a periodic amplification strategy. Instead of applying Grover search globally, amplitude amplification is integrated as a sparse amplification operator within the optimization loop. This "micro-dosage" strategy enables convergence acceleration while maintaining shallow circuit depth, thereby improving practical feasibility under NISQ constraints.   

Third, the hybrid optimization loop enhances robustness. As shown in Section IV, solution quality degrades gracefully under realistic noise conditions. This behavior arises from the interaction between shallow circuits, localized quantum operations, and measurement-driven feedback. From a distribution perspective, optimization is governed by iterative feedback dynamics rather than precise quantum state fidelity, which explains the observed robustness.

Taken together, these observations highlight a key design principle: practical quantum advantage in the NISQ era may emerge not from increasing quantum circuit complexity, but from aligning problem structure with available quantum resources.

\subsection{Limitations}
Despite its effectiveness, the proposed framework operates under a set of structural constraints that define its applicability and performance boundaries.

First, the current implementation relies on quantum simulation rather than physical quantum hardware. While the framework is designed to respect NISQ constraints, real hardware introduces additional challenges, such as gate errors, limited qubit connectivity, and calibration imperfections, which may affect performance.

Second, the amplitude amplification mechanism is implemented in a simplified and localized form. Instead of performing full Grover search with a global oracle, the framework adopts a periodic “micro-dosage” strategy. While effective in practice, this approach does not fully exploit the theoretical capabilities of amplitude amplification, reflecting a trade-off between quantum expressiveness and circuit feasibility.

Third, the effectiveness of the framework depends on the quality of problem decomposition. Although boundary refinement mitigates inconsistencies, clustering-based partitioning may introduce structural bias, particularly for irregular or non-uniform instances. In addition, the decomposition process may be sensitive to initialization, which can affect the stability of subproblem structures. This highlights a fundamental trade-off, where decomposition serves both as an enabler of scalability and a potential source of structural bias.

Fourth, the framework involves several hyperparameters, including subproblem size $M$ and amplification interval $\Delta G$. While empirical results suggest robustness, some parameters may require instance-dependent tuning, reflecting a trade-off between adaptability and parameter sensitivity.

Finally, the current evaluation does not include comparisons with state-of-the-art classical solvers such as Concorde or Lin–Kernighan–Helsgaun (LKH)\cite{1}. Incorporating such baselines would provide a more comprehensive assessment of the framework’s relative strengths and limitations.

Overall, these limitations indicate that the proposed framework is not intended to universally outperform classical methods, but to explore a practically feasible and structurally grounded hybrid optimization paradigm under NISQ constraints, where performance arises from the coordinated interaction between decomposition, stochastic search dynamics, and quantum enhancement.
\begin{table*}[htbp]
\centering
\caption{Resource statistics of the proposed framework on representative datasets.}
\label{tab:resource-stats}
\scriptsize

\begin{tabular}{lcccccc}
\toprule
Dataset & $n_{\text{cities}}$ & Subgroups & Max Size & Min Size & Max Qubits & Min Qubits \\
\midrule
Ulysses-16 & 16 & 5 & 7 & 1 & 21 & 1 \\
Ulysses-22 & 22 & 5 & 7 & 1 & 21 & 1 \\
Bays-29    & 29 & 5 & 8 & 3 & 24 & 6 \\
Att-48     & 48 & 8 & 8 & 3 & 24 & 6 \\
Eil-76     & 76 & 12 & 8 & 1 & 24 & 1 \\
Random-20  & 20 & 4 & 6 & 3 & 18 & 6 \\
Random-30  & 30 & 5 & 8 & 4 & 24 & 8 \\
Random-60  & 60 & 10 & 8 & 3 & 24 & 6 \\
Random-80  & 80 & 14 & 8 & 1 & 24 & 1 \\
Random-90  & 90 & 14 & 8 & 3 & 24 & 6 \\
Random-100 & 100 & 18 & 8 & 1 & 24 & 1 \\
\bottomrule
\end{tabular}

\medskip
\noindent \small $n_{\text{cities}}$: total number of cities in the instance.
Subgroups: number of final subclusters after $K$-means decomposition.
Max/Min Size: city count of the largest/smallest subcluster.
Max/Min Qubits: subcluster size $\times$ $\lceil\log_{2}$ subcluster size$\rceil$
of the largest/smallest subcluster.
\end{table*}
\subsection{Generalization and Design Principles}

Although this work focuses on the Traveling Salesman Problem, the proposed framework reflects a general hybrid optimization paradigm that can be extended to a broader class of combinatorial problems with similar structural characteristics. The key idea is to decompose a large-scale problem into resource-compatible subproblems, perform localized distribution-level exploration with periodic amplification, and reconstruct a globally consistent solution through classical refinement.

This structure is particularly suitable for problems that exhibit three structural properties: 
\begin{enumerate}
    \item decomposability, where the global problem can be partitioned into loosely coupled subcomponents.
    \item local evaluability, where subproblems can be independently optimized.
    \item  global consistency requirements, where local solutions must be integrated into a coherent structure.
\end{enumerate}
Representative examples include vehicle routing, scheduling and resource allocation, and large-scale network optimization.

Beyond specific applications, the framework suggests several general design principles for hybrid quantum–classical optimization under NISQ constraints. These principles reflect recurring patterns in the interaction between structural decomposition, distribution-level search, and localized enhancement.

Decomposition as a structural enabler.  Decomposition should be understood not merely as a computational simplification, but as a structural transformation that converts intractable global problems into tractable local subspaces. As shown in Section IV, scalability arises primarily from this structural reduction. However, decomposition alone is insufficient—methods that rely solely on clustering (e.g., QACO) fail to achieve comparable performance. This indicates that decomposition provides the foundation for scalability, but must be coupled with effective search dynamics.

Periodic amplification as a dynamic regulator.  Instead of applying quantum operations as standalone optimizers, the framework integrates amplitude amplification as a sparse amplification operator within the optimization loop. This "micro-dosage" strategy enables incremental improvement without introducing excessive circuit depth. The parameter study in Section IV reveals an inverted-U relationship between amplification frequency and solution quality, indicating that localized enhancement should be applied selectively and sparsely. Under NISQ constraints, effective quantum advantage arises not from maximizing circuit depth, but from regulating when and how quantum operations are introduced into the distribution evolution process.

Hybrid interaction as the source of performance.  The effectiveness of the framework arises from the interaction between classical and quantum components rather than from either component alone. Classical mechanisms provide reliable problem structuring and global refinement, while quantum-enhanced mechanisms improve exploration and convergence at the local level. As confirmed by the ablation study, neither component alone achieves optimal performance; instead, performance gains emerge from their coordinated interaction, forming a closed-loop optimization system.

From a broader perspective, these findings suggest that the role of quantum computation in near-term optimization is not to replace classical algorithms, but to function as a specialized component within a coordinated hybrid system\cite{wurtz2024solving}. The effectiveness of such systems depends on how structural decomposition, stochastic search dynamics, and quantum enhancement are jointly designed and integrated. This perspective suggests that the future of near-term quantum optimization may depend more on structured hybrid coordination than on isolated quantum acceleration alone.

Overall, effective hybrid optimization in the NISQ era is fundamentally structure-driven and interaction-centric, with performance emerging from the coordinated interaction between decomposition, distribution-level exploration, and periodic amplification.

\section{ Conclusion}
This work presents a NISQ-aware hybrid quantum–classical framework for scalable combinatorial optimization, with a focus on the Traveling Salesman Problem. By integrating structured decomposition, probabilistic quantum evolutionary search, and periodic amplitude amplification into a unified optimization pipeline, the proposed framework transforms an intractable global problem into a set of resource-compatible subproblems, enabling efficient optimization under realistic quantum hardware constraints. Extensive experiments on benchmark and synthetic datasets demonstrate consistent improvements in solution quality, scalability, and robustness, with performance gains becoming more pronounced as problem size increases. The results indicate that scalability is achieved through decomposition-driven control of computational complexity, while robustness is maintained under realistic NISQ noise conditions. Beyond empirical performance, this work highlights a general design paradigm for hybrid quantum–classical optimization under NISQ constraints. Within this paradigm, scalability emerges from resource-compatible decomposition, optimization efficiency from localized quantum enhancement, and solution quality from the coordinated interaction between probabilistic search and classical refinement. Overall, the results suggest that practical benefits in near-term quantum optimization may arise not from increasing circuit depth or relying on global quantum search, but from structuring optimization processes to align with available quantum resources. The proposed framework provides a concrete instance of this paradigm, offering a scalable and extensible pathway for applying hybrid quantum–classical methods to large-scale combinatorial problems under realistic hardware constraints. As quantum hardware continues to evolve, such structurally coordinated hybrid approaches may provide an effective transition pathway between current NISQ systems and future large-scale quantum optimization architectures.

\section*{Acknowledgment}

The authors acknowledge Yiwei Quantum Technology Co., Ltd. (Hefei 230088, China) for supporting this research and for providing the hybrid quantum–classical computing platform used in this study.

\bibliographystyle{IEEEtran}
\bibliography{references}

\EOD

\end{document}